\documentclass[aps,pra,preprint,groupedaddress,showpacs,%
tightenlines,
nofootinbib,floatfix]{revtex4}

\usepackage{amsmath,amsfonts,amssymb,bm}
\usepackage{graphicx}
\usepackage{dcolumn}
\usepackage{mathrsfs}

\renewcommand{\vec}[1]{\bm{\mathrm{#1}}} 

\begin{document}

\title{Differential cross sections for muonic atom scattering in 
       solid hydrogenic targets}

\author{Andrzej Adamczak}
\email{andrzej.adamczak@ifj.edu.pl}
\affiliation{Institute of Nuclear Physics, Polish Academy of Sciences, 
  PL-31342~Krak\'ow, Poland \\ 
  and Rzesz\'ow Technical University, PL-35959~Rzesz\'ow, Poland}

\date{\today}

\begin{abstract}
  The differential cross sections for low-energy muonic hydrogen atom
  scattering in solid molecular H$_2$, D$_2$ and T$_2$ targets under low
  pressure have been calculated for various temperatures. The
  polycrystalline fcc and hcp structure of the solid hydrogenic targets
  are considered. The Bragg and phonon scattering processes are
  described using the Debye model of a~solid. The calculated cross
  sections are used for Monte Carlo simulations of the muonic atom
  slowing down in these targets. They have been successfully applied for
  a~description of the production of the muonic atom beams in the
  multilayer hydrogenic crystals.
\end{abstract}


\maketitle


\section{Introduction}
\label{sec:intro}

The aim of this paper is to calculate the differential cross for muonic
hydrogen atom scattering in low-pressure molecular hydrogenic crystals
and to study the muonic atom deceleration in such crystals. The solid
hydrogenic targets have been used for the production of the muonic
hydrogen atom beams, which have been applied to the time-of-flight
measurements of various muonic-atom and muonic-molecular processes~(see
e.g., Refs.~\cite{know97,fuji00,porc01,mars01,wozn03,mulh06}). The bulk
hydrogenic crystals are employed in the investigations of muon-catalyzed
fusion of the hydrogen isotopes at high target densities and for various
populations of the rotational molecular
levels~\cite{demi96,peti01,mats03,toyo03,kawa03,ishi05,imao06}.
Hydrogenic solid targets are also used in a~novel method of spectroscopy
of the radioactive muonic atoms~\cite{stra04,stra05}.

Monte Carlo simulations of such experiments, which used the perfect-gas
model of a~target and the available differential cross
sections~\cite{brac89a,brac90} for the muonic atom scattering from
hydrogen-isotope nuclei, did not reproduce the experimental low-energy
($\lesssim{}1$~eV) data. Moreover, even the use of the cross sections
for scattering from isolated hydrogenic molecules~\cite{adam96a,adam06}
did not lead to an agreement between theory and
experiment~\cite{know97,peti01}. Thus, it is necessary to take into
account condensed-state effects of the muonic atom scattering in the
solids. In particular, interpreting the experimental data and planning
new experiments demand the knowledge of the differential cross sections
for scattering in the hydrogenic crystals.

A~few examples of the total cross sections for $p\mu$ and $d\mu$ atom
scattering in solid H$_2$ and D$_2$ at a~fixed target temperature
$T=3$~K were shown in~Ref.~\cite{adam99}. Below, a~method of calculation
of the partial differential cross sections for muonic hydrogen atom
scattering in the polycrystalline hydrogenic targets at various
temperatures is presented in detail. These cross sections are evaluated
using the amplitudes for muonic atom scattering from isolated hydrogenic
molecules~\cite{adam06} and the Van Hove response
function~$\mathcal{S}$~\cite{vanh54}. The isotropic Debye model of the
solid is employed.

In Sec.~\ref{sec:xsection_sol}, the coherent and incoherent cross
sections for the scattering in hydrogenic crystals are expressed by the
scattering amplitudes calculated for the isolated molecules. The
incoherent cross section, which includes the elastic scattering,
incoherent phonon scattering, and rotational-vibrational transitions in
the target molecules is discussed in~Sec.~\ref{sec:incoh_scatt}. The
Bragg scattering for the fcc and hcp structures, which are observed in
hydrogenic crystals, is considered in~Sec.~\ref{sec:Bragg}. The
inelastic coherent cross section, which leads to the creation or
annihilation of one phonon in a~muonic atom collision with the crystal
is calculated in~Sec.~\ref{sec:inel_coh_scatt}. Some examples of the
calculated cross sections for the homogeneous hydrogenic targets at
various temperatures are shown in Sec.~\ref{sec:xsec_examples}. Most
examples are given for the so-called ``normal''~\cite{soue86} targets
nH$_2$, nD$_2$, and~nT$_2$, which are characterized by the statistical
distribution of the molecular rotational levels $K=0$ and~1.

Symmetry of the wave function of a~homonuclear hydrogenic molecule is
definite. In the hydrogen case, the total nuclear spin~$I$ of H$_2$
equals 0 or~1. The singlet spin state $I=0$ is asymmetric. As a~result,
only the symmetric spatial wave functions with even values of the
rotational number~$K$ are allowed. The H$_2$ molecules in such states
are called parahydrogen molecules. For the symmetric triplet state
$I=1$, only odd values of~$K$ can occur. This defines the orthohydrogen
molecules. The situation is similar for the T$_2$ molecules, since the
spin~$s$ of proton or triton is equal to~1/2. In the deuterium case, one
has $s=1$. The total two-deuteron wave function must be symmetric.  For
the symmetric spin states $I=0$,~2, the spatial wave function is
symmetric, which corresponds to even~$K$. The D$_2$ molecules in such
states are called orthodeuterium molecules. The paradeuterium molecules
are characterized by $I=1$ and odd~$K$. The ortho and parastates of the
hydrogenic molecules are remarkably stable, in the absence of
a~catalyst~\cite{soue86}. As a~result, during rapid cooling of the
equilibrated nH$_2$, nD$_2$, or nT$_2$ gas, the even-$K$ states deexcite
to $K=0$ and the odd-$K$ states deexcite to $K=1$. The ortho-para
transitions can occur in certain collisions with muonic hydrogen atoms.
This is possible if the muonic atom interaction with a~homonuclear
molecule is spin-dependent. Such a~situation takes place when the
hydrogen isotope in the impinging atom is identical with those in the
homonuclear target molecule, e.g., in the $p\mu+\mathrm{H}_2$
scattering~\cite{brac89a,brac90}. In this case, the exchange forces
between the three identical nuclei should be taken into account. In
particular, they lead to a~high probability of the muon exchange between
two nuclei taking part in direct collision. When the spin projections of
these nuclei are opposite, the nuclear spin~$I$ of the target molecule
can change in the collision process. As a~result, the simultaneous
rotational ortho-para transition takes place. Such spin effects are
included in the presented calculations.

The obtained differential cross sections are used for Monte Carlo
simulations of the muonic atom deceleration. Some results of these
simulations are presented in Sec.~\ref{sec:mc_diff}.

\section{Coherent and incoherent cross sections}
\label{sec:xsection_sol}

A~muonic hydrogen atom~$a\mu$ can be approximately treated as a~small
neutron-like particle. Therefore, the methods derived for the
description of neutron scattering in condensed matter can be applied to
$a\mu$ scattering in dense hydrogenic targets. Below, the Van Hove
formalism~\cite{vanh54} is adapted to the calculation of the
differential cross sections for $a\mu$ scattering in the hydrogenic
crystals. In this formalism, the cross sections are expressed in terms
of the response function~$\mathcal{S}$, which depends solely on
properties of a~given target for fixed momentum and energy transfers.
A~definition of this function involves both quantum-mechanical and
statistical averaging over the target states at temperature~$T$.

It is assumed that there is no coupling between the translational and
collective motions of the molecules in the condensed target and the
molecular rotations and vibrations. The internal molecular degrees of
freedom are already included in a~single-molecule process~\cite{adam06}
and, therefore, do not enter into the response function. It is assumed
that a~bulk dense target is kept at a~sufficiently low pressure, so that
distortions of a~single bound molecule due to the interactions with
neighbors can be neglected. This assumption is fulfilled in the case of
low-pressure ($\ll{}10$~kbar) solid hydrogenic
targets~\cite{silv80,soue86}. The mean distance between the neighboring
molecules in such targets is several times greater than the diameter of
these molecules. The Van der Waals force between the molecules is weak.
As a~result, the rotational and vibrational numbers remain good quantum
numbers, although small broadening of certain excited molecular levels
takes place~\cite{vank83}. This broadening is not taken into account in
the presented calculations.

Low-pressure hydrogenic solids are quantum molecular crystals, which are
characterized by a~large amplitude of the zero-point vibrations of the
molecules in the lattice. The standard lattice dynamics can be applied
to these crystals, after certain renormalization of the
molecule-interaction potential~\cite{silv80,soue86}. Also, the Debye
model of a~solid can be used as a~reasonable approximation.

The wavelength of a~very slow ($\lesssim{}10$~meV) muonic hydrogen atom
is comparable to the nearest-molecule distance of
about~3.5~\AA\@~\cite{silv80,soue86}. Therefore, strong interference
effects can be observed at such energies. These effects are described
using a~conventional separation of the total differential cross sections
on the incoherent and coherent fractions. The coherent scattering takes
place only if specific geometrical conditions are fulfilled.  Analogous
to the neutron coherent scattering~\cite{love84}, the coherent cross
section for $a\mu$ scattering in a~solid single-isotope hydrogenic
target can be written down in the form:
\begin{equation}
  \label{eq:xcoh_def}
  \left(\frac{\partial^2\sigma}
    {\partial\varOmega\partial\varepsilon'}\right)_{\!\text{coh}} = 
  N_\text{mol}\, \frac{k'}{k}\, \sigma_\text{coh}\, 
  \mathcal{S}(\vec{\kappa},\omega) \,,
\end{equation}%
where $N_\text{mol}$ is the number of molecules in the target. The
energy transfer~$\omega$ and the momentum transfer~$\vec\kappa$ to the
lattice are, respectively, equal to
\begin{equation}
  \label{eq:transol_def}
  \omega = \varepsilon -\varepsilon' -\Delta E \,, \qquad\qquad
  \vec\kappa = \vec{k} - \vec{k}' ,
\end{equation}%
where $\varepsilon$ and $\varepsilon'$ denote the initial and final
kinetic energies of the scattered muonic atom and~$\Delta E$ is the
sum of the internal-energy changes of $a\mu$ and of the target
molecules. Vectors~$\vec{k}$ and~$\vec{k}'$ stand for the initial and
final momenta of~$a\mu$. These momenta and collision energies are
connected by the relations
\begin{equation}
  \label{eq:en_mom_rel}
  \varepsilon  = \tfrac{1}{2}k^2/M_{a\mu} \,,  \qquad
  \varepsilon' = \tfrac{1}{2}k'^2/M_{a\mu} \,,
\end{equation}
in which $M_{a\mu}$ denotes the $a_\mu$ mass. The
function~$\sigma_\text{coh}$ in Eq.~(\ref{eq:xcoh_def}) is expressed by
the amplitude~$\mathcal{F}^\text{mol}$ for $a\mu$ scattering from an
isolated molecule~\cite{adam06}
\begin{equation}
  \label{eq:amp_coh}
  \sigma_\text{coh} = \bigl| \overline{\mathcal{F}^\text{mol}} 
  \bigr|^2 .   
\end{equation}%
The horizontal bar stands here for averaging over a~random distribution
of the total spin~$\mathcal{J}$ of the $a\mu+$molecule system and over
a~distribution of the initial rotational states of the molecules. It is
assumed that there is no correlation between the direction of the
molecular spin~$\vec{I}$ and the lattice site.

Incoherent scattering does not include interference effects from
different molecules in the lattice. The incoherent cross section takes
a~general form~\cite{love84}
\begin{equation}
  \label{eq:xinc_def}
  \left(\frac{\partial^2\sigma}
    {\partial\varOmega\partial\varepsilon'}\right)_\text{inc} = 
  N_\text{mol}\, \frac{k'}{k}\, \sigma_\text{inc}\, 
  \mathcal{S}_i(\vec{\kappa},\omega) \,, 
\end{equation}%
where
\begin{equation}
  \label{eq:amp_incoh}
  \sigma_\text{inc} = \overline{\left|\mathcal{F}^\text{mol}\right|^2}
  -\bigl| \overline{\mathcal{F}^\text{mol}}\bigr|^2 ,   
\end{equation}%
and the incoherent response function
$\mathcal{S}_i(\vec{\kappa},\omega)$ is a~fraction of the total response
function~$\mathcal{S}(\vec{\kappa},\omega)$. In the limit of large
momentum transfers, the coherent processes disappear, so that
$\mathcal{S}(\vec{\kappa},\omega)\approx\mathcal{S}_i(\vec{\kappa},\omega)$.
The total differential cross section
$\partial^2\sigma/\partial\varOmega\partial\varepsilon'$ is a~sum of the
coherent~(\ref{eq:xcoh_def}) and incoherent~(\ref{eq:xinc_def}) cross
sections.

The inelastic scattering processes which change the internal state
of~$a\mu$ or that of the target molecule (such as the spin-flip and
isotopic-exchange reactions and the rotational-vibrational transitions)
are fully incoherent processes. No averaging over the states of the
different target molecules is performed and, therefore, in this case
$\sigma_\text{inc}$ reduces to the single-molecule squared amplitude
\begin{equation*}
  \sigma_\text{inc} = \left|\mathcal{F}^\text{mol}\right|^2  
  \quad \text{ and } \quad  \sigma_\text{coh} = 0 \,.
\end{equation*}

When the states of the muonic atom and molecule are not changed during
collision, $\sigma_\text{coh}$~and $\sigma_\text{inc}$ are equal to the
coherent and incoherent fractions of the elastic cross section for
$a\mu$ scattering from a~single molecule. Their values depend on a~given
choice of the hydrogen isotopes, the total spin~$F$ of~$a\mu$, the
population of the molecular rotational levels, and the collision energy.
In particular, when $\mathcal{F}^\text{mol}$ does not depend on the
spin~$\mathcal{J}$ and only one rotational state is populated, the
scattering is fully coherent
\begin{equation*}
  \sigma_\text{coh} =  
  \left|\overline{\mathcal{F}^\text{mol}}\right|^2 =
  \overline{\left|\mathcal{F}^\text{mol}\right|^2}  
  \quad \text{ and } \quad  \sigma_\text{inc} = 0 \,.
\end{equation*}
For example, such a~situation takes place in the case of elastic
scattering $d\mu+$H$_2$ in the ground rotational state $K=0$ of the
H$_2$ molecule.

In general, both $\sigma_\text{coh}$ and $\sigma_\text{inc}$ can have
nonzero values. In Table~\ref{tab:xmol_coh_incoh}, these functions are
shown in the limit $\varepsilon\to{}0$, for the cases
$p\mu+\mathrm{H}_2$, $t\mu+\mathrm{T}_2$, and $d\mu+\mathrm{D}_2$ and
for several values of~$F$ and~$K$.
\begin{table}[htb]
  \begin{center}
    \caption{Coherent~$\sigma_\text{coh}$ and
      incoherent~$\sigma_\text{inc}$ elastic cross sections (in
      $10^{-20}$~cm$^2$/sr) for a~single hydrogenic molecule, at the
      collision energy~$\varepsilon\to{}0$.
      \label{tab:xmol_coh_incoh}}
    \begin{ruledtabular}  
    \newcolumntype{.}{D{.}{.}{3.3}}
    \begin{tabular}{c c c . .}
      \multicolumn{1}{c}{process}&
      \multicolumn{1}{c}{$F$}&
      \multicolumn{1}{c}{$K$}&
      \multicolumn{1}{c}{$\sigma_\text{coh}$}&
      \multicolumn{1}{c}{$\sigma_\text{inc}$} \\
      \hline
      &  0  &  0  &  88.5   &   0.0    \\
      $p\mu+$H$_2$
      &  0  &  1  &  88.7   &   0.0    \\
      &  1  &  0  & 126.2   &   0.0    \\
      &  1  &  1  & 124.8   &  87.6    \\
      \hline 
      &  0  &  0  & 325.7   &   0.0    \\
      $t\mu+$T$_2$
      &  0  &  1  & 324.9   &   0.0    \\
      &  1  &  0  & 342.9   &   0.0    \\
      &  1  &  1  & 342.4   &  10.5    \\
      \hline
      &  1/2  &  0  & 64.7   &  0.15   \\
      $d\mu+$D$_2$
      &  1/2  &  1  & 64.4   &  0.06   \\
      &  3/2  &  0  & 64.5   &  0.38   \\
      &  3/2  &  1  & 64.3   &  0.15   \\
    \end{tabular}  
   \end{ruledtabular}
  \end{center}
\end{table}

\section{Incoherent scattering}
\label{sec:incoh_scatt}

The incoherent response function~$\mathcal{S}_i$ can be rigorously
calculated for harmonic crystals~\cite{vanh54,sing60}. Using the
so-called phonon expansion of~$\mathcal{S}_i$ for a~cubic Bravais
lattice (one molecule per lattice cell), the incoherent cross
section~(\ref{eq:xinc_def}) takes the form:
\begin{equation}
  \label{eq:xinc_phon_exp}
  \begin{split}
     \left(\frac{\partial^2\sigma}
     {\partial\varOmega\partial\varepsilon'}\right)_{\!\text{incoh}} = ~& 
     N_\text{mol}\, \frac{k'}{k}\, \sigma_\text{inc}\, \exp(-2W) \\
     &\times \left[ \delta(\omega) +\sum_{n=1}^{\infty} g_n(\omega) 
     \, \frac{(2W)^n}{n !} \right],
  \end{split} 
\end{equation}
where $\exp(-2W)$ denotes the Debye-Waller factor, which is familiar in
the theory of neutron scattering. This expansion is also a~fair
approximation for any cubic lattice, e.g., the fcc structure. The
exponent of the Debye-Waller factor is by definition equal to
\begin{equation}
  \label{eq:2W_def}
  2W(\vec{\kappa})\equiv\left\langle(\vec{\kappa}\cdot\vec{u})^2
  \right\rangle_T \,,
\end{equation}
in which $\vec{u}$ is the displacement of the molecule from its lattice
site and $\langle\ldots\rangle_T$ denotes the quantum-mechanical and
statistical averaging at temperature~$T$. In particular, for
a~cubic-crystal structure
\begin{equation}
  \label{eq:2W_u}
  2W =\tfrac{1}{3}\langle \vec{u}^2 \rangle_T \, \kappa^2 \,.
\end{equation}
It follows from~(\ref{eq:2W_def}) that $2W$ does not disappear at $T=0$
because the mean square displacement tends in this limit to a~finite
value determined by the zero-point vibrations of the molecule in the
lattice. The functions~$g_n$ in Eq.~(\ref{eq:xinc_phon_exp}) are defined
as
\begin{equation}
  \label{eq:g_n}
  \begin{split}
    & g_1(w) = \frac{1}{2W}\,\frac{\kappa^2}{2M_\text{mol}} 
    \frac{Z(w)}{w} \left[\,n_\text{B}(w)+1\right] \,, \\[3pt]
    & g_n(w) = \int_{-\infty}^{\infty} \text{d} w' \,
    g_1(w-w')\, g_{n-1}(w') \,, \\[3pt]
    & \int_{-\infty}^{\infty}\text{d} w\, g_n(w)  = 1 \,,
  \end{split}
\end{equation}
where $M_\text{mol}$ is the mass of the molecule. The normalized
density of vibrational states~$Z(w)$ in the isotropic Debye model has
the following form:
\begin{equation}
  \label{eq:Z_Debye}
  Z(w)\equiv
  \begin{cases}
    3\, w^2/w_\text{D}^3 & \text{~if~} \,
    w \leq w_\text{D} \\
    0 & \text{~if~} \, w > w_\text{D} \,,
  \end{cases}
\end{equation}%
where $w_\text{D}=k_\text{B}\varTheta_\text{D}$ is the Debye energy
corresponding to the Debye temperature~$\varTheta_\text{D}$
($k_\text{B}$~stands for the Boltzmann constant). For the low-pressure
hydrogenic crystals, $\varTheta_\text{D}$~is on the order of~100~K.
Although differences between the masses of the hydrogen isotopes are
quite large, isotopic effects in~$\varTheta_\text{D}$ for these crystals
are quite small, due to quantum effects~\cite{silv80,soue86}. The
function $n_\text{B}(w)$ in~Eqs.~(\ref{eq:g_n}) denotes the Bose factor
\begin{equation}
  \label{eq:Bose_fac}   
  n_\text{B}(w) = \left[\,\exp(w/k_\text{B}T)-1\right]^{-1},
\end{equation}%
which determines the phonon population at a~given~$T$.

Note that there is no momentum-conservation condition in the incoherent
cross section~(\ref{eq:xinc_phon_exp}). The $\delta$~function in
Eq.~(\ref{eq:xinc_phon_exp}) represents the recoil-less scattering from
the rigid lattice. In the case of elastic scattering
($\Delta{}E=\nobreak{}0$), $a\mu$ cannot change its energy in this
process. This is connected with a~large mass of the solid target. The
next terms give broad distributions corresponding to the subsequent
multiphonon processes. In particular, the term with~$n=1$ describes
incoherent $a\mu$ scattering with simultaneous creation or annihilation
of one phonon. The amplitudes of all the processes are proportional to
the Debye-Waller factor, which decreases with the rising momentum
transfer.

Elastic $a\mu$ scattering from a~free molecule is described by a~single
function. In a~solid, this function is replaced by a~set of functions
corresponding to the strictly elastic and multiphonon processes which
are proportional to~$\kappa^{2n}$. The same conclusion can be drawn for
any incoherent process, e.g., for a~rotational excitation of the target
molecule. In such a~scattering event, the nonphonon cross section is
connected with the energy transfer~$\Delta{}E$ to the molecule.  The
higher energy transfers are due to the rotational excitation with
simultaneous phonon creation.

When the momentum transfer is small ($2W\ll{}1$), only the several
lowest terms in Eq.~(\ref{eq:xinc_phon_exp}) are important. On the other
hand, in the limit $2W\gg 1$ (weak binding), many multiphonon terms give
comparable contributions to the cross section. For sufficiently
large~$\kappa^2$, it is convenient to use the impulse approximation in
which $\mathcal{S}_i$ takes a~general Gaussian form for any target
described by a~time-independent Hamiltonian~\cite{vine58}. In this
approach, the incoherent cross section is equal to
\begin{equation}
  \label{eq:xinc_asym}
  \begin{split}
     \left(\frac{\partial^2\sigma}
     {\partial\varOmega\partial\varepsilon'}\right)_\text{inc} = &~
     N_\text{mol}\, \frac{k'}{k}\, \sigma_\text{inc} \\   
     &\times \frac{1}{\Delta_\text{R} \sqrt{\,\pi}} \,
     \exp\left[-\left(\frac{\omega - \omega_\text{R}}
    {\Delta_\text{R}} \right)^{\!2} \,\right],
  \end{split}
\end{equation}
where 
\begin{equation}
  \label{eq:Dopp_width}
  \Delta_\text{R} = \sqrt{8\,\mathscr{E}_T\, \omega_\text{R}/3}  
\end{equation}
is the Doppler width of the asymptotic form of~$\mathcal{S}_i$. The mean
kinetic energy of a~single molecule in the lattice at temperature~$T$ is
denoted by~$\mathscr{E}_T$ and $\omega_\text{R}$ is the recoil energy
\begin{equation}
  \label{eq:recoil_mol}
  \omega_\text{R} = \tfrac{1}{2}\kappa^2/M_\text{mol} \,.
\end{equation}%
For a~harmonic solid
\begin{equation}
  \label{eq:meankin}
  \mathscr{E}_T = \tfrac{3}{2} \int_0^{\infty} \text{d} w \,
  Z(w)\,w \left[n_\text{B}(w)+\tfrac{1}{2}\right].
\end{equation}
This energy contains a~contribution from the zero-point vibrations of
the lattice molecules. In the case of low-pressure hydrogenic crystal,
one has $T/\varTheta_\text{D}\ll{}1$. In this limit, $\mathscr{E}_T$~and
the effective target temperature~$T_\text{eff}$
\begin{equation}
  \label{eq:T_eff}
  T_{\text{eff}} = \tfrac{2}{3}\mathscr{E}_T/k_\text{B}\,.
\end{equation}
are well approximated by  
\begin{equation}
  \label{eq:2W_asym}
  \mathscr{E}_T = \tfrac{9}{16}\, w_\text{D} \,, \quad
  \text{ and } \quad
  T_{\text{eff}} = \tfrac{3}{8}\, \varTheta_\text{D}\,, 
\end{equation}
respectively. In particular, for a~3-K zero-pressure solid deuterium
target with $\varTheta_\text{D}=108$~K, one obtains
$\mathscr{E}_T=5.2$~meV and $T_{\text{eff}}\approx{}40$~K. Thus,
$T_{\text{eff}}\gg{}T$, which means that the cross sections calculated
for 3-K perfect-gas~D$_2$ should not be used for a~description of the
solid D$_2$ target at the same temperature. Relatively high values
of~$\mathscr{E}_T$ in the solid H$_2$ and~D$_2$ targets have been
experimentally confirmed using deep inelastic neutron
scattering~\cite{herw90,momp96,zopp01}.

At very high collision energies $\varepsilon\gg{}w_\text{D}$ and
$\varepsilon\gg\Delta{}E$, the approximation
$\varepsilon'\approx\varepsilon$ ($k'\approx{}k$) is valid. As a~result,
the cross section~(\ref{eq:xinc_asym}) averaged over~$\varepsilon'$
tends to the static approximation
\begin{equation}
  \label{eq:xinc_static}
  \left(\frac{\text{d}\sigma}{\text{d}\varOmega}\right)_\text{inc} = 
  N_\text{mol}\,\sigma_\text{inc} \,. 
\end{equation}
This cross section no longer depends on the target structure and is
equal to the sum of the free-molecule incoherent cross sections. Since,
at high energies, coherent effects disappear, the total differential
cross section $\text{d}\sigma/\text{d}\varOmega$ is given by
Eq.~(\ref{eq:xinc_static}) with $\sigma_\text{inc}$ replaced
by~$|\mathcal{F}^\text{mol}|^2$.

\begin{figure}[htb]
  \begin{center}
    \includegraphics[width=7cm]{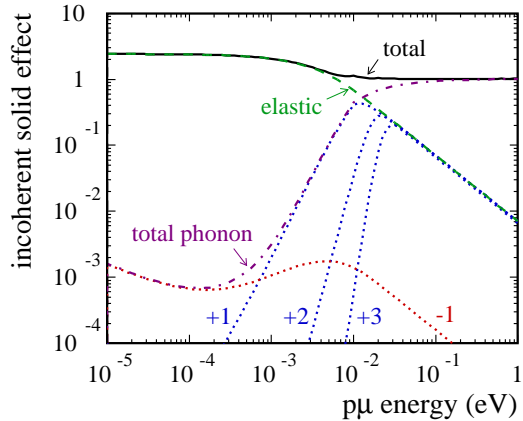}
    \caption{(Color online) Function $\mathcal{C}_\text{inc}$ versus
      energy~$\varepsilon$ for incoherent elastic scattering of
      $p\mu$-atom in 3-K solid-hydrogen. Subsequent
      phonon-creation processes are denoted by~``$+1$'' (one-phonon),
      ``$+2$'' (two-phonon), and~``$+3$'' (three-phonon). Label~``$-1$''
      stands for annihilation of one phonon.
      \label{fig:C_sol_el}}
  \end{center}
\end{figure}

The influence of the lattice binding on the incoherent scattering is
described by the following function:
\begin{equation}
  \label{eq:C_sol}
  \mathcal{C}_\text{inc} \equiv \frac{1}{4\pi N_\text{mol}}\, 
  \left(\frac{M_{a\mu}}{\mathcal{M}}\right)^{\! 2} \!
  \int \text{d}\varOmega \text{d}\varepsilon' \, 
  \left(\frac{\partial^2\sigma}
  {\partial\varOmega\partial\varepsilon'}\right)_\text{inc},  
\end{equation}
where $\mathcal{M}$ denotes the reduced mass of the $a\mu+$molecule
system. It is assumed here that $\mathcal{F}^\text{mol}$ is constant.
The function~$\mathcal{C}_\text{inc}$ for elastic ($\Delta{}E=0$) $p\mu$
scattering in 3-K solid~H$_2$ at zero pressure is presented
in~Fig.~\ref{fig:C_sol_el}.
The different curves show contributions from the nonphonon, phonon
creation, and phonon annihilation processes. Every single process falls
exponentially for sufficiently large energies, which is due to the
Debye-Waller factor in~Eq.~(\ref{eq:xinc_phon_exp}). At~energies below
a~few~meV, the nonphonon elastic scattering is dominant. The phonon
annihilation, which is strongly suppressed by the Bose factor at~3~K,
prevails over the phonon creation only at
$\varepsilon\lesssim{}0.7$~meV. The subsequent phonon-creation
processes appear when $\varepsilon$ rises. Finally, they dominate the
scattering at~$\varepsilon\gg{}w_\text{D}$, where the
static-approximation limit is reached and $\mathcal{C}_\text{inc}=1$.
For $\varepsilon\to{}0\,$, one has
$\mathcal{C}_\text{inc}=(M_{a\mu}/\mathcal{M})^2=2.24$.
\begin{figure}[htb]
  \begin{center}
    \includegraphics[width=7cm]{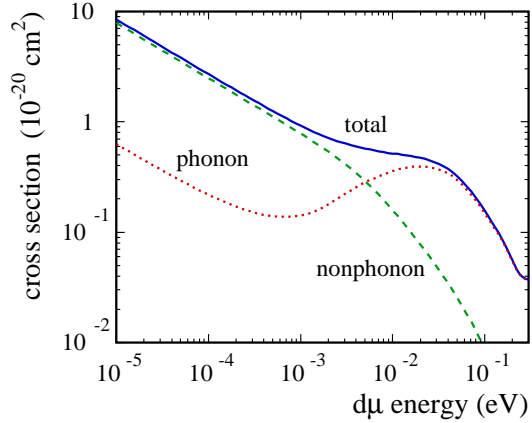}
    \caption{(Color online) Phonon and nonphonon contributions to 
      the cross section for the rotational deexcitation
      $d\mu(F=3/2)$+D$_2(K=1)\to{}d\mu(F=3/2)$ 
      +D$_2(K=0)$ in solid deuterium at $T=3$~K.
      \label{fig:rot_fon}}
  \end{center}
\end{figure}
In~Fig.~\ref{fig:rot_fon}, contributions of the nonphonon and phonon
fractions of~$\mathcal{S}_i$ to the total cross section are plotted for
the rotational deexcitation $K=1\to{}0$ of a~D$_2$~molecule bound in
3-K solid deuterium, in collision with a~$d\mu(F=3/2)$~atom. The cross
section is proportional to~$\varepsilon^{-1/2}$ at $\varepsilon\to{}0$.
Even at the lowest energies, the phonon processes are more important
than in the elastic case (cf.~Fig.~\ref{fig:C_sol_el}) because the
rotational energy release of~7.5~meV is comparable to
$w_\text{D}\approx{}9$~meV.

\section{Coherent elastic scattering}
\label{sec:Bragg}

The fraction of the response function~$\mathcal{S}$ that describes
coherent elastic scattering (Bragg scattering) is well known in the
neutron theory (see e.g., Ref.~\cite{love84}). When applied
to~Eq.~(\ref{eq:xcoh_def}), it leads to the following coherent elastic
cross section for a~perfect-crystal lattice with $N_d$ identical
molecules per unit cell of volume~$V_0$:
\begin{equation}
  \label{eq:xBragg_nb}
  \begin{split}
    \left(\frac{\text{d}\sigma}{\text{d}\varOmega}\right)_{\!\text{coh}}^{\!\text{el}} 
    = &~N\,\frac{(2\pi)^3}{V_0}
    \sum_{\vec{\tau}}|F_N(\vec{\tau})|^2\\
    &\times\delta(\vec{\kappa} - \vec{\tau}) \,
    \exp\left[-2W(\kappa^2)\right],
  \end{split}
\end{equation}
$F_N(\vec{\tau})$ being the unit-cell structure factor
\begin{equation}
  \label{eq:cell_fac}
  |F_N(\vec{\tau})|^2 = \sigma_\text{coh} \,
  \Bigl|\sum_{d=1}^{N_d} \exp(i\vec{\tau}\cdot\vec{d})\Bigr|^2.
\end{equation}
The summation in Eq.~(\ref{eq:xBragg_nb}) is performed over the
reciprocal-lattice vectors~$\vec{\tau}$. The vector $\vec{d}$ denotes
the position of a~given molecule in the unit cell and $N$~is the number
of the unit cells in the crystal. The Bragg scattering of~$a\mu$ in
a~large crystal is possible only when the momentum transfer is equal to
one of the reciprocal lattice vectors
\begin{equation}
  \label{eq:cond_Bragg_kappa}
  \vec{\kappa} = \vec{k} - \vec{k}' = \vec{\tau} \,.
\end{equation}
The intensity of scattering peaks is determined by the Debye-Waller
factor and by the value of~$\sigma_\text{coh}$ for the specific
momentum~$k$ and the scattering angle~$\vartheta$. At high energies, the
elastic Bragg scattering vanishes. Below the Bragg cutoff energy
\begin{equation}
  \label{eq:e_Bragg-cutoff}
  \varepsilon_\text{B} = \frac{1}{8}\tau_\text{min}^2/M_{a\mu}\,,
\end{equation}%
the condition~(\ref{eq:cond_Bragg_kappa}) cannot possibly be fulfilled
and the elastic coherent scattering disappears. The value of the
shortest nonzero vector~$\vec{\tau}$ is denoted here
by~$\tau_\text{min}$.

The Bragg-cutoff energies for various combinations of the muonic atoms
and hydrogenic targets are shown in Table~\ref{tab:brag_cutoff}, for the
3-K fcc~targets at zero-pressure.  They have been calculated using the
hydrogenic-crystal data from Refs.~\cite{soue86,silv80}.
\begin{table}[htb]
  \begin{center}
    \caption{The Bragg cutoff energy~$\varepsilon_{_\text{B}}$ (in meV) for 
      $p\mu$, $d\mu$, and $t\mu$ atom scattering in the 3-K fcc crystals
      of~nH$_2$, nD$_2$, and nT$_2$ at zero pressure.
      \label{tab:brag_cutoff}}
    \begin{ruledtabular}
    \newcolumntype{.}{D{.}{.}{4.4}}
    \begin{tabular}{c . . .}
      \multicolumn{1}{l}{lattice} &
      \multicolumn{1}{c}{~~nH$_2$}  &
      \multicolumn{1}{c}{~~nD$_2$}  &
      \multicolumn{1}{c}{~~nT$_2$}  \\
      \hline
      $p\mu$ & 1.94 & 2.13 & 2.21 \\
      $d\mu$ & 1.02 & 1.12 & 1.17 \\
      $t\mu$ & 0.69 & 0.76 & 0.79 \\
    \end{tabular}
   \end{ruledtabular}    
  \end{center}
\end{table}

When scattering takes place from a~polycrystalline sample, which is
usually the case in the muonic-hydrogen physics,
Eq.~(\ref{eq:xBragg_nb}) can be averaged over all orientations of the
lattice. This gives the following cross section:
\begin{equation}
  \label{eq:xBragg_poly}
  \begin{split}
    \left(\frac{\text{d}\sigma}{\text{d}\varOmega}\right)%
    _{\!\text{coh}}^{\!\text{el}} & = 
    N\,\frac{2\pi^2}{V_0}\,\frac{1}{k^2}\,
    \sum_{\vec{\tau}} |F_N(\vec{\tau})|^2 \, \frac{1}{\tau} \\ 
    &\times\delta\!\left(1-\frac{\tau^2}{2k^2}-\cos\vartheta\right)
    \exp\left[-2W(\kappa^2)\right].
  \end{split}
\end{equation}
The scattering now takes place in the Debye-Scherrer cones around the
direction of~$\vec{k}$. The cones have the semi-angles~$\vartheta$
subject to the condition $\cos\vartheta=1-\tau^2/2k^2$.
\begin{figure}[htb]
  \begin{center}
    \includegraphics[width=7cm]{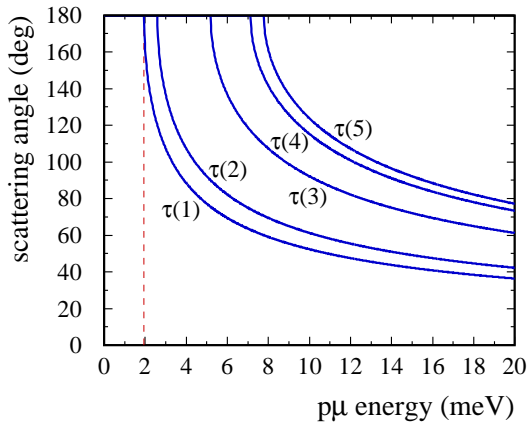}
    \caption{(Color online) Scattering angle~$\vartheta$ in Bragg 
      scattering of $p\mu$ in 3-K~nH$_2$ versus the collision energy 
      and the value of the reciprocal-lattice vector~$\vec{\tau}$. 
      The fcc polycrystalline structure of the target is assumed.
      \label{fig:angbrag_ppp}}
  \end{center}
\end{figure}
In~Fig.~\ref{fig:angbrag_ppp}, $\vartheta$~is plotted as a~function of
the $p\mu$ kinetic energy in a~3-K nH$_2$ target with the
polycrystalline fcc structure. The scattering angle is shown for the
five smallest values $\tau(i),\, i=1,\ldots,5$ of the
vectors~$\vec{\tau}$. As one can see from Eq.~(\ref{eq:xBragg_poly}) and
Fig.~\ref{fig:angbrag_ppp}, the subsequent Bragg peaks appear at
$\vartheta=180^{\circ}$. When $\varepsilon$ increases, this backward
scattering opens into a~cone which moves continuously towards the
forward direction. The total cross
section~$\sigma_\text{coh}^\text{el}(\tau)$ is expressed by the formula
\begin{equation}
  \label{eq:xBragg_tot}
  \sigma_\text{coh}^\text{el} =  N \, 
  \frac{4\pi^3}{V_0}\, \frac{1}{k^2} \, 
  \sum_{\vec{\tau}}^{\tau<2k} |F_\tau|^2\, \frac{1}{\tau}\, 
  \exp\left[-2W(\tau^2)\right] \,.
\end{equation}
where $z(\tau)$ is the number of $\vec{\tau}$-vectors with the same
magnitude~$\tau$ and $|F_\tau|^2$ denotes the average value
of~$|F_N(\vec{\tau})|^2$ for just these vectors.

\begin{figure}[htb]
  \begin{center}
     \includegraphics[width=7cm]{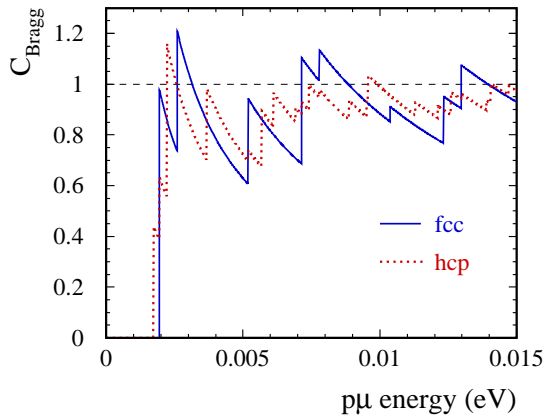}
     \caption{(Color online) The Bragg pattern in the case of scattering 
       of a~$p\mu$ atom in a~3-K polycrystalline nH$_2$ target with the 
       fcc and hcp structures.
       \label{fig:bfac_hf}}
  \end{center}
\end{figure}

In order to illustrate the energy dependence of interference effects,
the function
\begin{equation}
  \label{eq:C_Bragg}
  \mathcal{C}_\text{Bragg}(\varepsilon) 
  \equiv \frac{4\pi^3}{V_0}\,
  \frac{|F_\tau|^2}{N_d\, \sigma_\text{coh}}\, \frac{1}{k^2} \, 
  \sum_{\vec{\tau}}^{\tau<2k} \frac{1}{\tau} \,,
\end{equation}
is plotted in~Fig.~\ref{fig:bfac_hf} for $p\mu$ scattering in
polycrystalline nH$_2$ with the fcc and hcp structures. This function
is equal to the total Bragg cross section for scattering in a~rigid
lattice [$\exp(-2W)\equiv{}1$], calculated per one molecule. It is
assumed that $\sigma_\text{coh}$ is constant.
The coherent scattering is forbidden below
$\varepsilon_\text{B}\approx{}2$~meV. The subsequent Bragg peaks appear
for the rising $p\mu$ energy. The Bragg-cutoff energy is slightly lower
for the hcp structure.

\section{Coherent inelastic scattering}
\label{sec:inel_coh_scatt}

The inelastic coherent scattering is connected with the energy transfer
between $a\mu$ and the collective degrees of freedom of the target. Such
a~scattering describes the coherent annihilation and creation of
phonons, but it does not include changes of the internal state of the
target molecule ($\Delta{}E=0$). Thus, the energy transfer to the
lattice is $\omega=\varepsilon-\varepsilon'$. The inelastic coherent
cross section for $a\mu$ scattering in a~hydrogenic solid can be
calculated using the methods developed for the coherent scattering of
neutrons~\cite{love84}. In this paper, coherent effects are taken into
account only in the most important one-phonon processes. An estimation
of the multiphonon cross sections is much more difficult. Therefore,
these cross sections are calculated in the incoherent approximation. In
the case of a~Bravais harmonic lattice, the coherent one-phonon cross
section is given by the following formula:
\begin{equation}
  \label{eq:xcoh_phon}
  \begin{split}
    \left(\frac{\partial^2\sigma}
      {\partial\varOmega\partial\varepsilon'}
    \right)^\text{inel}_\text{coh,1} & =\frac{k'}{k}\, 
    \frac{(2\pi)^3}{V_0}\, \frac{1}{2M_\text{mol}} 
    \sum_{\vec{\tau}} \sigma_\text{coh}\\
    &\times\exp\left[-2W(\vec{\kappa})\right] \sum_{j,\vec{q}}
    \frac{\lvert\vec{\kappa}\cdot\vec{\sigma}^j\rvert^2}{w_j} \\ 
    &\times 
    \Bigl[n_\text{B}^j(\vec{q})\,\delta\bigl(\omega+w_j\bigr)\,
      \delta(\vec{\kappa}+\vec{q}-\vec{\tau}) \\
      &+\bigl(n_\text{B}^j(\vec{q})+1\bigr) \,
      \delta\bigl(\omega-w_j\bigr)\,
      \delta(\vec{\kappa}-\vec{q}-\vec{\tau})\Bigr],
  \end{split}
\end{equation}
$\vec{q}$ being the phonon momentum. There are three phonon-polarization
vectors~$\vec{\sigma}^j(\vec{q})$ ($j=1,2,3$) with the corresponding
phonon energies~$w_j(\vec{q})$. The Bose factor for fixed
$\vec{\omega}_j$ and~$\vec{q}$ is denoted by~$n_\text{B}^j(\vec{q})$.
The dispersion relation $w_j=w_j(\vec{q})$ can be obtained by solving
the lattice dynamics and is often measured in experiments, e.g., by
means of neutron scattering.  Because of the translational symmetry of
the lattice, $\vec{\omega}_j$~is a~periodic function of~$\vec{q}$
\begin{equation}
  \label{eq:dispersion_period}
  w_j(\vec{q}) = w_j(\vec{q}+\vec{\tau}) \,.
\end{equation}
Thus, one can restrict values of~$\vec{q}$ to the first Brillouin
zone~($\vec{\tau}=\vec{0}$) in order to know~$\vec{\omega}_j$ at
any~$\vec{q}$. The dispersion relation for small~$q$ takes the form
\begin{equation}
  \label{eq:dispersion_rel}
  w_j(\vec{q}) = c_\text{s}(\vec{\sigma}^j)\, q \,,
\end{equation}
where $c_\text{s}(\vec{\sigma}^j)$ is the sound velocity in a~given
crystal.

The cross section~(\ref{eq:xcoh_phon}) consists of the two terms. The
first, which contains the expression %
$\delta(\omega+w_j)\delta(\vec{\kappa}+\vec{q}-\vec{\tau})$ describes
annihilation of one phonon. The term with the factor %
$\delta(\omega-w_j)\delta(\vec{\kappa}-\vec{q}-\vec{\tau})$ corresponds
to $a\mu$ scattering with simultaneous one-phonon creation.  The
annihilation processes vanish when the lattice temperature approaches
zero, since there are no phonons at~$T=0$. The $\delta$~functions in
Eq.~(\ref{eq:xcoh_phon}) represent conservation of both the energy and
momentum, which is a~basic feature of the coherent inelastic scattering.
By virtue of Eqs.~(\ref{eq:transol_def}) and~(\ref{eq:xcoh_phon}), the
initial and final energies and momenta of the muonic atom fulfill the
conditions
\begin{equation}
  \label{eq:cond_annih}
  \varepsilon' = \varepsilon + w_j \,, \qquad
  \vec{k}' = \vec{k} + \vec{q} -\vec{\tau}\,,
\end{equation}
in the case of one-phonon annihilation and obey the conditions
\begin{equation}
  \label{eq:cond_creat}
  \varepsilon' = \varepsilon - w_j \,, \qquad
  \vec{k}' = \vec{k} - \vec{q} -\vec{\tau}\,,
\end{equation}
when one phonon is created. Therefore, for a~fixed scattering angle,
only phonons with a~specific~$\vec{q}$ and~$w_j(\vec{q})$ lead to the
coherent one-phonon scattering.

It can be shown that the coherent one-phonon creation near the forward
direction $\vec{k}'\approx\vec{k}$ (possible when $\vec{\tau}=\vec{0}$)
takes place only if the velocity of the impinging atom is greater than
the sound velocity~$c_\text{s}$~\cite{love84}. This means that the
slowing down of the muonic atom via this process is impossible at the
lowest energies.

The isotropic Debye model of a~solid is now used for the coherent
phonon scattering. The average sound velocity~$c_\text{s}$ is substituted
in~Eq.~(\ref{eq:dispersion_rel}), which is taken as the first
approximation to a~real-crystal dispersion law. This assumption is
reasonable because the coherent phonon processes are important in muonic
atom deceleration only at the lowest energies. At higher energies, 
incoherent scattering prevails. The factor %
$\lvert\vec{\kappa}\cdot\vec{\sigma}^j(\vec{q})\rvert^2$ in 
Eq.~(\ref{eq:xcoh_phon}) is replaced by its average value over
a~surface with a~fixed~$w$. Such an average for the cubic crystals 
equals $\kappa^2/3$, which is a~fair approximation even for most
noncubic crystals~\cite{love84}. Using the definition
\begin{equation}
  \label{eq:vib_density}
  Z(w) \equiv \frac{1}{3N_\text{mol}}\sum_{j,\vec{q}}
    \delta\bigl[w-w_j(\vec{q})\bigr] 
\end{equation}
of the normalized density of vibrational states, the summation over
$j$ and $\vec{q}$ in~Eq.~(\ref{eq:xcoh_phon}) is replaced by the
integration
\begin{equation*}
  \sum_{j,\vec{q}} \longrightarrow 3 N_\text{mol} 
  \int_0^{w_\text{D}} \text{d} w \, Z(w)\,.
\end{equation*}
The result of this integration is then averaged over the directions
of~$\vec{\tau}$ and~$\vec{q}$. Finally, one obtains the following
formula:
\begin{equation}
  \label{eq:xcoh_phon_av}
  \begin{split}
    \left(\frac{\partial^2\sigma}
      {\partial\varOmega\partial\varepsilon'}
    \right)^\text{inel}_\text{coh,1} = &~N_\text{mol}\, \frac{k'}{k}\, 
    \sigma_\text{coh}\, \exp(-2W)\,\frac{\kappa^2}{2M_\text{mol}}\\ 
    &\times  \, Z(\omega)\,
    \frac{n_\text{B}}{(\omega)+1}{\omega}\, \mathcal{R}(\kappa,\omega) \,,
  \end{split}
\end{equation}
where 
\begin{equation*}
  \begin{split}
    \mathcal{R}(\kappa,\omega)  = &~\frac{\pi^2}{V_0}\, 
    \frac{1}{\kappa{}q}\, \Theta(w_\text{D}-|\omega|)\\ 
    &\times \sum_{\vec{\tau}} \frac{1}{\tau}\,
    \Theta(\tau-|\kappa-q|)\, \Theta(\kappa+q-\tau) 
  \end{split}
\end{equation*}
and $\Theta$ denotes the Heavyside function. There is a~direct relation
between this cross section and the incoherent one-phonon cross
section
\begin{equation}
  \label{eq:xcoh_xincoh}
  \left(\frac{\partial^2\sigma}
    {\partial\varOmega\partial\varepsilon'}
  \right)^{\!\text{inel}}_{\!\text{coh,1}} = 
  \frac{\sigma_\text{coh}}{\sigma_\text{inc}} 
  \,\mathcal{R}(\kappa,\omega)\,
  \left(\frac{\partial^2\sigma}
    {\partial\varOmega\partial\varepsilon'}
  \right)_\text{inc,1}. 
\end{equation}
In the limit $\omega\to{}0$, the function~$\mathcal{R}$ is proportional
to the factor~$\delta(\kappa-\tau)$. This gives a~geometrical condition
similar to that for Bragg scattering in a~polycrystalline target. Thus,
in this limit, the coherent one-phonon scattering displays the same
pattern of scattering peaks as that observed in the
$a\mu$-Bragg-scattering case (apart from~$\tau=0$).

To describe the coherent phonon effects, the function
\begin{equation}
  \label{eq:C_sol_coh}
  \mathcal{C}_\text{coh} \equiv \frac{1}{4\pi N_\text{mol}}\,
  \left(\frac{M_{a\mu}}{\mathcal{M}}\right)^{\! 2} \!
  \int \text{d}\varOmega  \text{d}\varepsilon' \, 
  \left(\frac{\partial^2\sigma}{\partial\varOmega\partial\varepsilon'}
  \right)^\text{inel}_\text{coh,1}
\end{equation}
is defined. A~constant $\mathcal{F}^{mol}$ is assumed here. 
\begin{figure}[htb]
  \begin{center}
    \includegraphics[width=7cm]{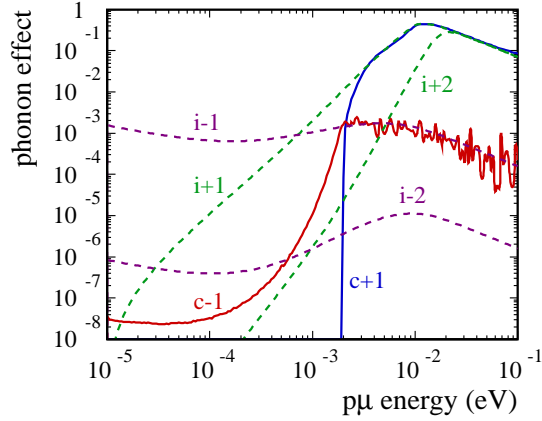}
    \caption{(Color online) Functions $\mathcal{C}_\text{coh}$ and
      $\mathcal{C}_\text{inc}$ for $p\mu$ scattering in the fcc
      polycrystalline~H$_2$ at~3~K. Coherent one-phonon annihilation
      (``c$-1$'') and creation (``c$+1$'') are shown using the solid
      lines. Incoherent (dashed lines) one-phonon processes have the
      labels: ``i$-1$'' (annihilation) and ``i$+1$'' (creation). The
      labels ``i$-2$'' and ``i$+2$'' stand for the two-phonon incoherent
      processes.
      \label{fig:foncoh_ppp_3k}}
  \end{center}
\end{figure}
\begin{figure}[htb]
  \begin{center}
      \includegraphics[width=7cm]{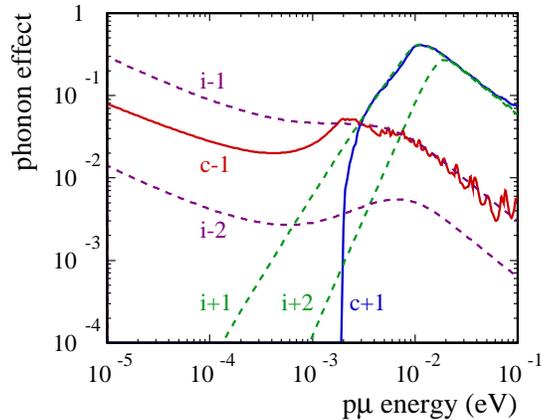}
      \caption{(Color online) The same as in 
        Fig.~\ref{fig:foncoh_ppp_3k}, for $T=13.9$~K.
        \label{fig:foncoh_ppp_14k}}
  \end{center}
\end{figure}
In~Figs.~\ref{fig:foncoh_ppp_3k} and~\ref{fig:foncoh_ppp_14k}, both the
coherent~$\mathcal{C}_\text{coh}$ and
incoherent~$\mathcal{C}_\text{inc}$ functions are shown in the case of
$p\mu$ scattering in the fcc polycrystalline~H$_2$ at $T=3$ and~13.9~K.
The one-phonon coherent and incoherent functions differ strongly below
the Bragg-cutoff energy, especially at the lowest temperatures.  On the
other hand, the total coherent cross sections approach the corresponding
incoherent ones at $\varepsilon$~greater than a~few~meV.  Since the
energy spectrum of a~created phonon is broad ($\approx{}9$~meV), the
total cross section for one-phonon coherent creation is smooth.
Oscillations of the coherent-annihilation cross section apparent at
higher energies are due to a~narrow width ($\approx$~1~meV) of the
energy spectrum of phonons which exist in the low-temperature targets.
The functions $\mathcal{C}_\text{inc}(\varepsilon)$, which describe the
incoherent annihilation and creation of two phonons, are also plotted in
Figs.~\ref{fig:foncoh_ppp_3k} and~\ref{fig:foncoh_ppp_14k}. One can see
that the incoherent approximation to multiphonon processes is reasonable
because $\mathcal{C}_\text{coh}$~for one-phonon coherent annihilation is
much stronger than the correction from two-phonon incoherent
annihilation.  The only exception can be seen at~$\varepsilon\ll{}1$~meV
in the 3-K target.  However, this energy interval is not important when
the slowing down of the $p\mu$-atom from $\varepsilon\sim{}1$~eV is
considered.

\section{Examples of cross sections for homogeneous hydrogenic targets}
\label{sec:xsec_examples}

Some examples of the cross sections calculated for solid H$_2$, D$_2$,
and T$_2$ targets used in the TRIUMF, JINR, and RIKEN-RAL experiments
are shown below. They include all the processes which preserve the total
spin~$F$ of the muonic atom. In all the cases, the polycrystalline
disordered (no specific orientations of molecules in the $K=1$ state)
fcc structure has been assumed. At temperatures higher than 4~K, the
targets can have the polycrystalline hcp structure, which depends on the
method of preparation and history of a~given
target~\cite{silv80,soue86}. However, the cross sections for the fcc
lattice are good approximations for the hcp structure (apart from the
Bragg cross section, which is calculated separately for the hcp case).
The reason for this is that the first three shells of neighbor molecules
in both lattices are identical if the anisotropic interactions and
orientations of molecules are neglected.  The molar volumes of both
structures are the same. The coherent phonon scattering is important
below a~few meV, where the acoustic-phonon
approximation~(\ref{eq:dispersion_rel}) is valid for the two structures,
with practically the same value of the mean sound velocity.

Although the Debye temperatures found in the literature are measured or
calculated for crystals that do not exactly correspond to the targets
used in muonic-hydrogen physics, these values are taken as a~reasonable
approximation. The inaccuracy involved by such an approach is estimated
to be below~10\%. When, for a~specific condition, there is no data
available, the Debye energy is calculated using the following equation:
\begin{equation}
  \label{eq:edeb-sound}
  w_\text{D} = 2(3\pi^2)^{1/3}\, c_\text{s}/a_\text{fcc} \,,
\end{equation}
with the sound velocity $c_\text{s}$ and the lattice
constant~$a_\text{fcc}$ taken or calculated from Ref.~\cite{soue86}.

\begin{figure}[htb]
  \begin{center}
    \includegraphics[width=7cm]{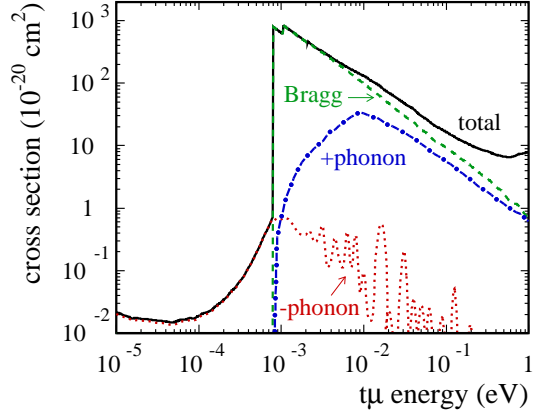}
    \caption{(Color online) Total cross section for $t\mu(F=0)$ 
      scattering in zero-pressure solid nT$_2$ at $T=3$~K. 
      Contributions from Bragg scattering, phonon 
      creation (+phonon), and phonon annihilation ($-$phonon) 
      processes are also shown.
      \label{fig:xfttt11_3k}}
  \end{center}
\end{figure}

\begin{figure}[htb]
  \begin{center}
    \includegraphics[width=7cm]{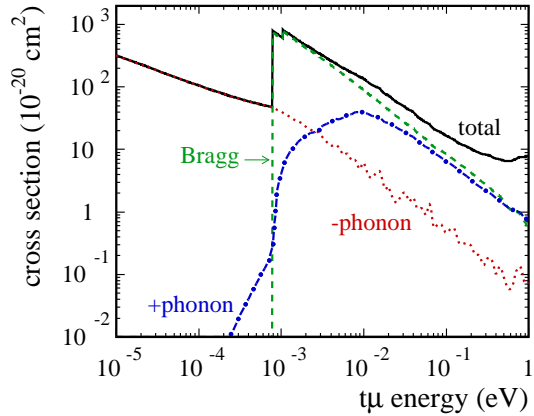}
    \caption{(Color online) The same as in~Fig.~\ref{fig:xfttt11_3k} 
      for $t\mu(F=0)$ and $T=20$~K.
      \label{fig:xfttt11_20k}}
  \end{center}
\end{figure}

\begin{figure}[htb]
  \begin{center}
    \includegraphics[width=7cm]{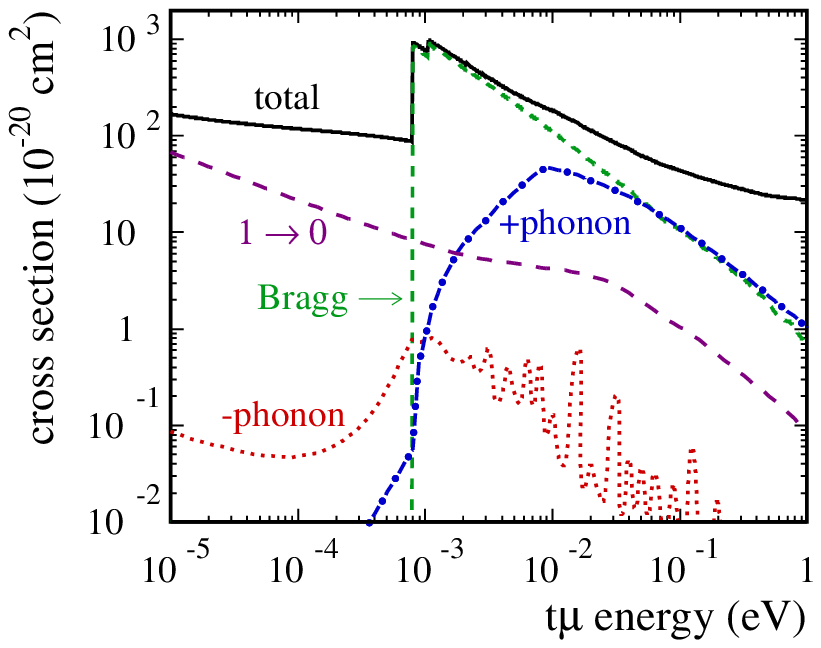}
    \caption{(Color online) The same as in~Fig.~\ref{fig:xfttt11_3k} 
      for $t\mu(F=1)$ and $T=3$~K. The label ``1$\to$0'' denotes 
      the rotational deexcitation $K=1\to{}0$ of the target molecule.
      \label{fig:xfttt22_3k}}
  \end{center}
\end{figure}

\begin{figure}[htb]
  \begin{center}
    \includegraphics[width=7cm]{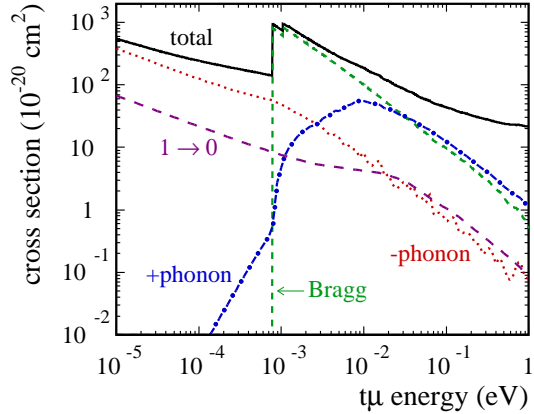}
    \caption{(Color online) The same as in~Fig.~\ref{fig:xfttt22_3k} 
      for $t\mu(F=1)$ and~$T=20$~K.
      \label{fig:xfttt22_20k}}
  \end{center}
\end{figure}

\begin{figure}[htb]
  \begin{center}
    \includegraphics[width=7cm]{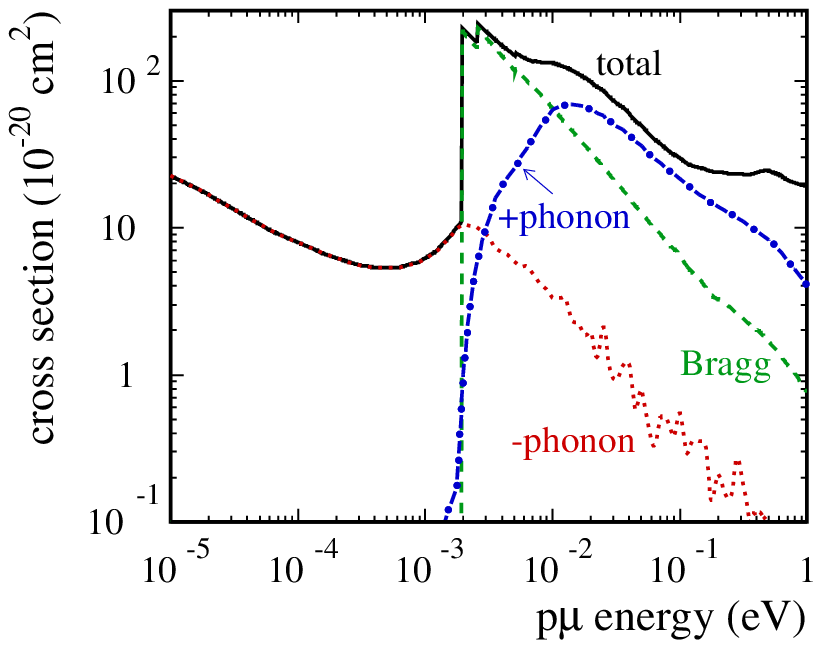}
    \caption{(Color online) Total cross section for $p\mu(F=0)$ 
      scattering in zero-pressure solid nH$_2$ at $T=13.9$~K. 
      The notation is the same as in~Fig.~\ref{fig:xfttt11_3k}. 
      \label{fig:xfppp11_14k}}
  \end{center}
\end{figure}

\begin{figure}[htb]
  \begin{center}
    \includegraphics[width=7cm]{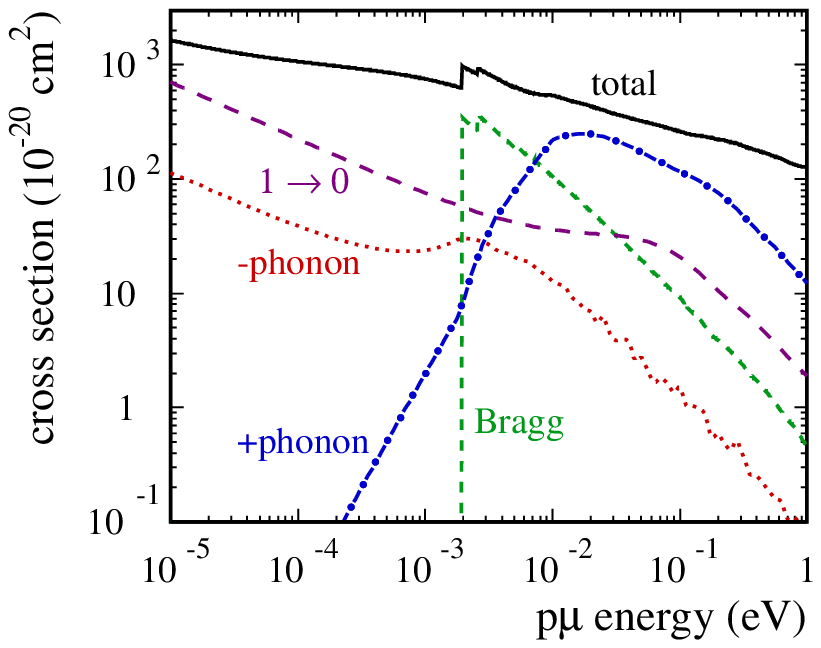}
    \caption{(Color online) Total cross section for $p\mu(F=1)$ 
      scattering in zero-pressure solid nH$_2$ for $F=0$ at
      $T=13.9$~K. The notation is the same as in~Fig.~\ref{fig:xfttt22_3k}. 
      \label{fig:xfppp22_14k}}
  \end{center}
\end{figure}

\begin{figure}[htb]
  \begin{center}
    \includegraphics[width=7cm]{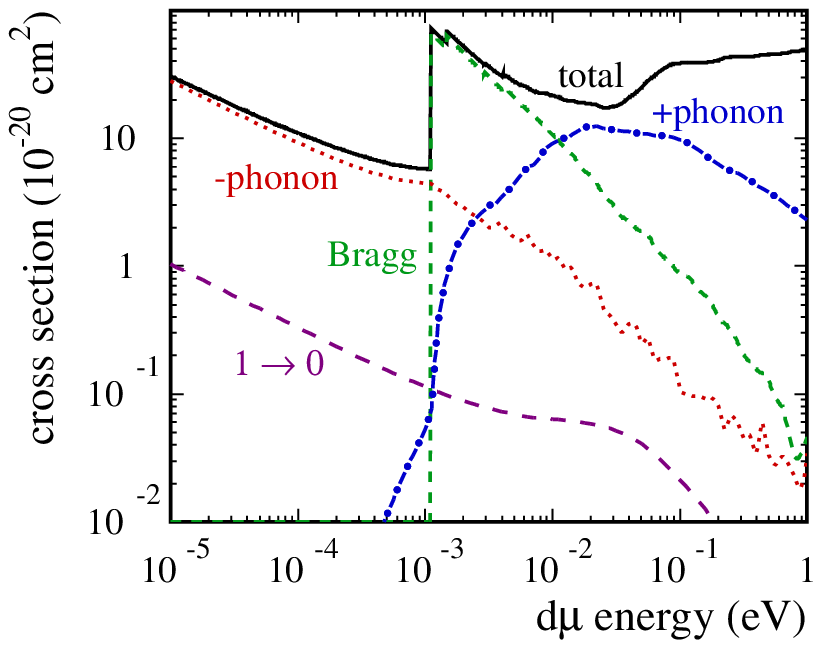}
    \caption{(Color online) Total cross section for $d\mu(1/2)$ scattering 
      in zero-pressure solid nD$_2$ at $T=18.7$~K. The notation is 
      the same as in~Fig.~\ref{fig:xfttt22_3k}.
      \label{fig:xfddd11_19k}}
  \end{center}
\end{figure}

In Figs.~\ref{fig:xfttt11_3k}--\ref{fig:xfddd11_19k}, the label
``total'' and the solid lines denote the total cross sections, which
include all the coherent and incoherent processes.  Contributions to the
total cross section from the Bragg scattering (short-dashed line),
phonon annihilation (dotted line, label ``$-$phonon''), phonon creation
(dash-dotted line, label $+$phonon''), and the rotational transition
$K=1\to{}0$ (long-dashed line) are also plotted. The lines which
describe the phonon processes do not include the rotational-vibrational
transitions, although all the incoherent inelastic processes can take
place with simultaneous phonon annihilation or creation. Thus, the
rotational cross sections shown in these figures also contain the phonon
terms. All the presented cross sections are calculated per one
hydrogenic molecule bound in the solid target. The cross sections for
$t\mu$ atom scattering in zero-pressure solid~nT$_2$ at~$T=3$ and 20~K
(just below the melting point) are shown
in~Figs.~\ref{fig:xfttt11_3k}--\ref{fig:xfttt22_20k}.
For the ground state $F=0$ of the $t\mu$ spin, the scattering at the
lowest energies is almost fully coherent (a~small incoherence is caused
by the slightly different single-molecule scattering amplitudes for the
two lowest rotational states of~T$_2$). Below the Bragg cutoff, the
elastic and phonon-creation coherent processes are forbidden. Thus, in
this energy region, the coherent annihilation and incoherent phonon
scattering are the only inelastic processes. At the lowest temperatures,
however, these processes are strongly suppressed by the Bose factor. As
a~result, the total cross section in Fig.~\ref{fig:xfttt11_3k} falls by
many orders of magnitude below~$\varepsilon_\text{B}$. This leads to
a~large increase of the $a\mu$ mean free path, which results in an
enhanced emission of the cold ($\sim{}1$~meV) muonic atoms from the thin
solid targets. This phenomenon has been already observed in the TRIUMF
experiment~\cite{wozn03}, in the case of $p\mu$ atoms. When the target
temperature is raised, the low-energy phonon processes become more
important and, near the melting temperature, the phonon cross section is
quite large at $\varepsilon\to{}0$ (see Fig.~\ref{fig:xfttt11_20k}).

For $F=1$, both the coherent and incoherent processes are significant
since $\sigma_\text{inc}$ is appreciable
(cf.~Table~\ref{tab:xmol_coh_incoh}). Therefore, the total cross section
below~$\varepsilon_{_\text{B}}$ does not fall so greatly, compared to
the $F=0$ case. The other important difference between the $F=0$ and
$F=1$ cases is the rotational deexcitation $K=1\to{}0$, which is due to
the muon exchange between the tritium nuclei with the opposite spin
projections. At the lowest energies, such a~transition leads to an
effective acceleration of the muonic atom. This transition is forbidden
when the total spin of $t\mu(F=0)$ is conserved in the collision
process.  On the other hand, such a~transition can take place with the
simultaneous excitation $F=0\to{}1$ of the $t\mu$ spin state. However,
the spin-flip process has a~threshold of 0.237~eV (in the $t\mu+t$
center of mass) so that this reaction cannot occur at the lowest
energies.

Similar features of the cross sections appear in the $p\mu$ atom
scattering in zero-pressure solid nH$_2$. Figures~\ref{fig:xfppp11_14k}
and~\ref{fig:xfppp22_14k} show the cross sections for $F=0$ and $F=1$
at~$T=13.9$~K (below the melting temperature of the solid~H$_2$).  The
analogous cross sections for $T=3$~K were presented
in~Ref.~\cite{adam99,wozn03}. For $p\mu(F=1)$, the incoherent scattering
at the lowest energies is more pronounced than in the tritium case since
$\sigma_\text{inc}$ is comparable to~$\sigma_\text{coh}$
(cf.~Table~\ref{tab:xmol_coh_incoh}).

\begin{figure}[htb]
  \begin{center}
    \includegraphics[width=7cm]{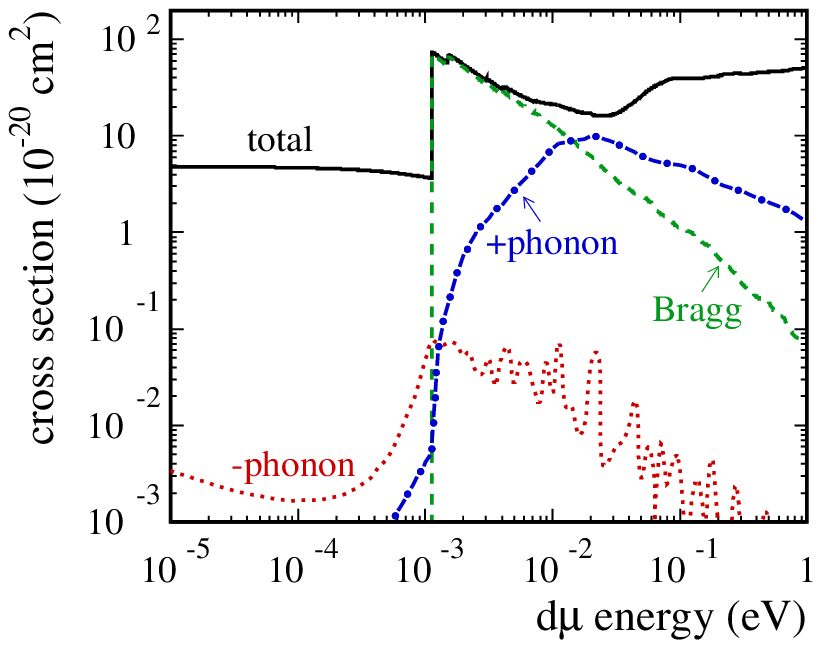}
    \caption{(Color online) Total cross section for $d\mu(3/2)$ scattering 
      in zero-pressure solid ortho-D$_2$ at $T=3$~K. The notation is 
      the same as in~Fig.~\ref{fig:xfttt11_3k}.
      \label{fig:xfddd22k0_3k}}
  \end{center}
\end{figure}
In the deuterium case, $d\mu$ atom scattering from the D$_2$ molecule
does not significantly depend on the $d\mu$-atom total
spin~\cite{adam96a}. Therefore, the cross sections for $d\mu(F=1/2)$ and
$d\mu(F=3/2)$ scattering in solid~D$_2$ are very similar. The rotational
deexcitation $K=1\to{}0$ of the target molecule in $d\mu$ scattering
which conserves the total spin~$F$ is allowed for both the $F=1/2$ and
$F=3/2$ states. In Fig.~\ref{fig:xfddd11_19k}, the total cross section
for $d\mu(F=1/2)$ scattering in zero-pressure nD$_2$ is plotted for the
target temperature just below the melting point. For the ortho-D$_2$
target, there is no rotational deexcitation. Thus, in this case, the
phonon annihilation is the only $d\mu$-acceleration process, which is
shown in~Fig.~\ref{fig:xfddd22k0_3k} for $T=3$~K.

\begin{figure}[htb]
  \begin{center}
    \includegraphics[width=7cm]{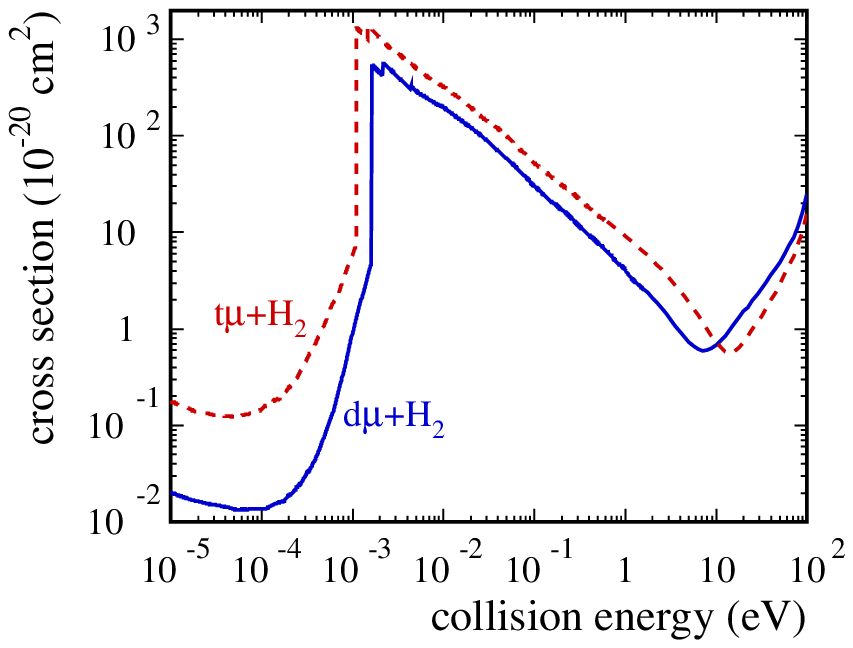}
    \caption{(Color online) Total cross sections for $d\mu$ and $t\mu$ 
      scattering in 3-K solid nH$_2$.
      \label{fig:xdtppel_3k}}
  \end{center}
\end{figure}
In~Fig.~\ref{fig:xdtppel_3k}, the total cross sections for $d\mu$ and
$t\mu$ scattering in 3-K solid~nH$_2$ are plotted. They can also be
applied for solid H$_2$ with a~very small admixture of D$_2$ or~T$_2$.
When a~$\mu^{-}$ beam is stopped in such a~target, it is mainly $p\mu$
atoms which are created. Some of them can then hit a~D$_2$ or a~T$_2$
molecule, which mostly leads to the isotopic muon
transfer~\cite{adam92}. As a~result, the released $d\mu$ or $t\mu$ atom
has a~kinetic energy on the order of a~few tens~eV. During the
deceleration process, these atoms are emitted to vacuum while they reach
the Ramsauer-Townsend minima, which is apparent at about~10~eV. This
mechanism has been used for the production of the energetic (1-10~eV)
$d\mu$ and $t\mu$ beams
at~TRIUMF~\cite{know97,fuji00,porc01,mars01,wozn03,mulh06}. The
calculated differential cross sections establish the basis for accurate
simulations of the emission of such muonic-atom beams.
Fig.~\ref{fig:xdtppel_3k} shows also that one should expect an enhanced
emission of the cold ($\sim{}1$-meV) muonic atoms from the target, due
to the falloff of the total cross sections below the Bragg cutoff.

\section{Muonic atom deceleration in hydrogenic crystals}
\label{sec:mc_diff}

When a~$\mu^{-}$ beam is stopped in a~hydrogenic target, the muonic
hydrogen atoms are created in the highly excited Coulombic states.
During the deexcitation process, these atoms gain kinetic energy.  As
a~result, the ground-state 1S~atoms have a~broad energy distribution
(see e.g., Refs.~\cite{mark94,abbo97,faif01,jens02,pohl06}), which can
even extend to~$10^3$~eV. Since the muonic atomic and muonic-molecular
processes strongly depend on the $a\mu$ energy, it is necessary to know
its time evolution.  In the solid targets, the $a\mu$ deceleration from
the highest energies to about~0.1~eV is very fast ($\sim{}1$~ns).
Therefore, only the last stage of $a\mu$ slowing down, when solid-state
effects are important, is considered here. The Maxwell distribution of
the initial $a\mu$ kinetic energy with the mean
value~$\varepsilon_\text{avg}=1$~eV and the statistical population of
the initial $a\mu$ spin states are assumed. The mean
energy~$\varepsilon_\text{avg}$ as a~function of time and temperature
has been evaluated using the calculated differential cross sections and
Monte Carlo simulations.

\begin{figure}[htb]
  \begin{center}
    \includegraphics[width=7cm]{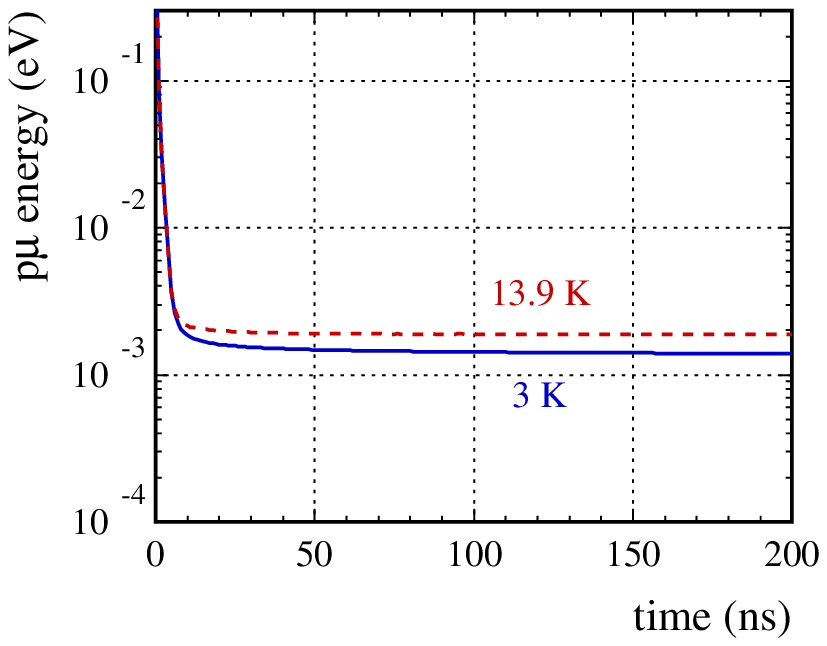}
    \caption{(Color online) Mean energy of a~$p\mu$ atom in solid nH$_2$ 
     versus time at $T=3$ and~13.9~K. 
     \label{fig:aven_p}}
  \end{center}
\end{figure}
\begin{figure}[htb]
  \begin{center}
    \includegraphics[width=7cm]{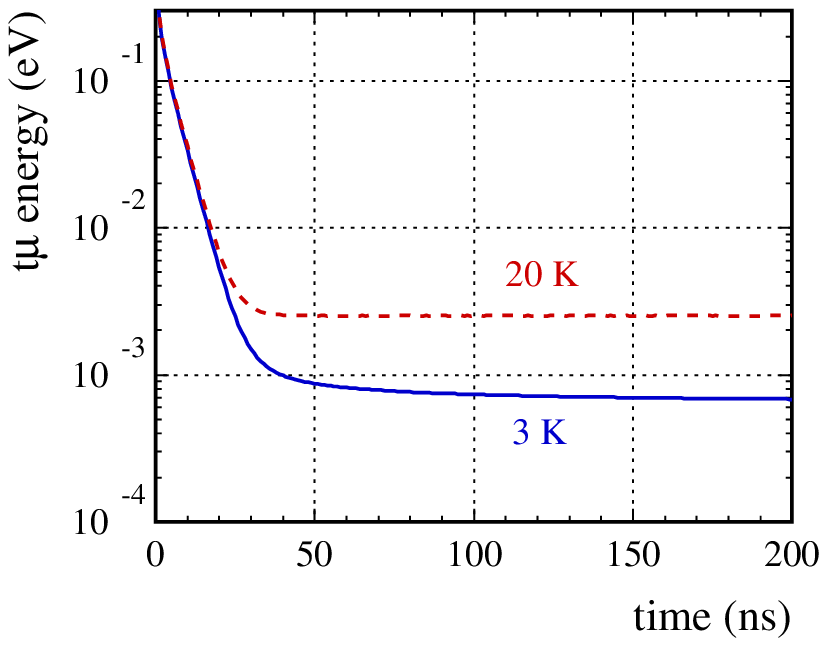}
    \caption{(Color online) Mean energy of a~$t\mu$ atom in solid nT$_2$ 
      versus time at $T=3$ and~20~K. 
      \label{fig:aven_t}}
  \end{center}
\end{figure}
In~Figs.~\ref{fig:aven_p}--\ref{fig:aven_t}, $\varepsilon_\text{avg}$
for the $p\mu$ atom in solid~nH$_2$ and for the $t\mu$ atom in
solid~nT$_2$ is plotted versus time. This energy is additionally
averaged over the muonic-atom spin~$F$. In fact, after about~1~ns, the
higher spin states $F=1$ of the $p\mu$ and $t\mu$ atoms are depleted
because of the large downwards spin-flip cross section~\cite{brac89a}.
The deceleration of~$p\mu$'s to~$\varepsilon_\text{avg}=10$~meV is very
fast. This stage is much longer for~$t\mu$'s as the cross section
$t\mu(F=0)+t$ is relatively small at the lowest energies~\cite{brac90}.
Below~$10$~meV, solid-state effects are very important and, therefore,
they strongly affect the slowing down. This is especially visible
at~3~K, when the phonon processes are suppressed and the classical
thermal energy $3k_\text{B}T/2$ is much smaller than the Bragg cutoff
energy~$\varepsilon_\text{B}$ (see Fig.~\ref{fig:xfttt11_3k}).
Below~$\varepsilon_\text{B}$, there is no effective deceleration
mechanism. As a~result, at the lowest temperatures, the mean energy of
the muonic atoms in the steady state is much higher than the thermal
energy. This energy is achieved after about 200~ns. The deceleration
below~10~meV is slow since the scattering is dominated by the elastic
Bragg process. On the other hand, when the target temperature tends to
the melting point, $3k_\text{B}T/2$ is greater than
$\varepsilon_\text{B}$ and the deceleration is faster. Also, the
steady-state~$\varepsilon_\text{avg}$ already approaches~$\mathscr{E}_T$
(cf.~the cross sections from Figs.~\ref{fig:xfttt11_20k}
and~\ref{fig:xfppp11_14k}). One sees that the difference between the
steady-state values of~$\varepsilon_\text{avg}$ is much smaller than the
difference of the corresponding classical thermal energies for the two
limiting temperatures. This is especially pronounced in solid~H$_2$,
where $\varepsilon_\text{B}$~is the greatest and the melting temperature
is the smallest, compared to the solid D$_2$ and T$_2$ targets.
\begin{figure}[htb] 
  \begin{center}
    \includegraphics[width=7cm]{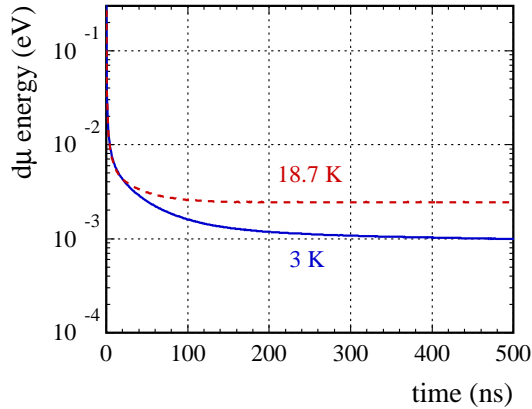}
    \caption{(Color online) Mean energy of a~$d\mu$ atom in solid 
      nD$_2$ versus time at~$T=3$~and 18.7~K.
      \label{fig:aven_d}}
  \end{center}
\end{figure}
The analogous deceleration functions for $d\mu$ atom in solid~nD$_2$ are
plotted in~Fig.~\ref{fig:aven_d}. Since the spin-flip rate for this case
is very small~\cite{brac89a}, the deexcitation of the upper spin state
is completed only after about~100~ns. This additionally extends the
slowing down process as the hyperfine-splitting energy of~48.5~meV is
released.

\begin{figure}[htb]
  \begin{center}
    \includegraphics[width=7cm]{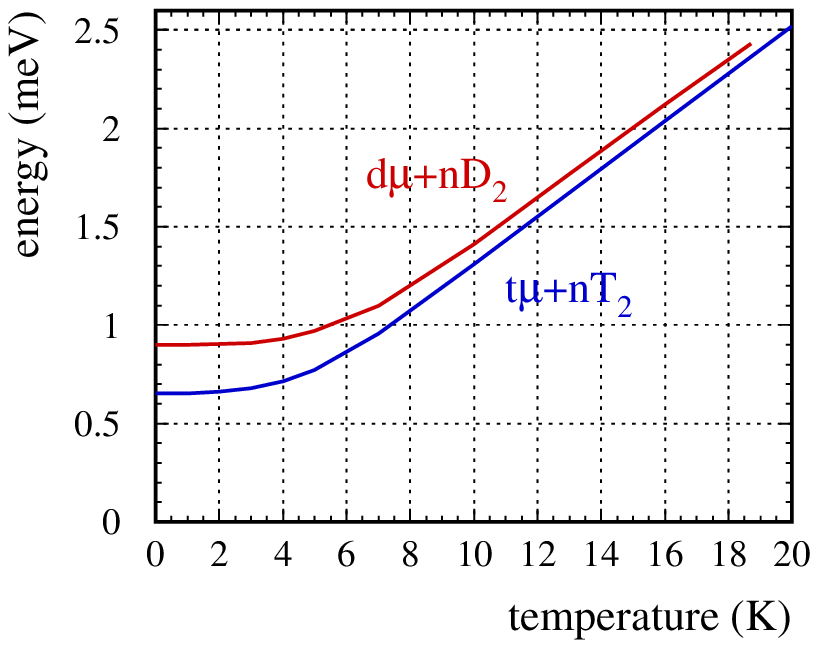}
    \caption{(Color online) The calculated mean energy of a~$d\mu$ atom
      in solid nD$_2$ and of a~$t\mu$ atom in solid nT$_2$ at large times 
      versus the target temperature. 
      \label{fig:dt_entemp}}
  \end{center}
\end{figure}
In~Fig.~\ref{fig:dt_entemp}, the dependence of~$\varepsilon_\text{avg}$
in the steady state is shown as a~function of the target temperature,
for the solid~nD$_2$ and~nT$_2$ targets. A~full set of the differential
cross sections has been calculated for about ten values of the
temperature. Then, the mean energy has been estimated by means of
Monte-Carlo simulations. Below a~few Kelvin, $\varepsilon_\text{avg}$~is
almost constant, which is due to the lack of effective deceleration
mechanisms at $\varepsilon\lesssim{}1$~meV. At higher temperatures,
$\varepsilon_{\text{avg}}$~changes almost linearly and approaches
$3k_\text{B}T/2$ near the melting point.

\section{Conclusions}

The differential cross sections for muonic hydrogen atom scattering in
the polycrystalline hydrogenic targets have been calculated using the
scattering amplitudes for single molecules and the response function for
the isotropic solid.  A~certain number of important approximations has
been made in the presented calculations. The scattering amplitudes for
isolated hydrogenic molecules are obtained using the corresponding
amplitudes for muonic atom scattering from free hydrogen-isotope
nuclei~\cite{brac89a,brac90} and the first-Born approximation. It is
assumed that the molecular vibrations are harmonic and that there is no
coupling between the vibrational and rotational degrees of freedom. The
presence of electrons in the target molecules is taken into account by
introduction of the effective electron-screening potential. These
approximations have been discussed in detail in Ref.~\cite{adam06}. In
the case of solid molecular targets, it is assumed that the
rotational-vibrational structure of the bound molecules is not changed,
which is a~fair approximation for the low-pressure hydrogenic
crystals~\cite{soue86,silv80}. These targets are described in the
harmonic approximation and the response function for the isotropic Debye
solid is used. The Debye model is applied for estimation of the coherent
and incoherent phonon cross sections. Coherent and incoherent effects
are separately taken into account only in the most important one-phonon
scattering. The cross sections for multiphonon processes are calculated
in the incoherent approach.

It happens that the muonic atom coherent scattering is very important in
the H$_2$, D$_2$, and~T$_2$ targets at $\varepsilon\lesssim{}10$~meV. In
particular, a~rapid decrease of the total cross section below the Bragg
cutoff energy leads to an enhanced emission of cold muonic atoms at
temperatures $T\lesssim{}5$~K, where the phonon annihilation processes
are strongly suppressed. The enhanced emission of $p\mu$ atoms has been
observed and is well described by the calculated differential cross
sections~\cite{wozn03}. The coherent phonon creation disappears below
about 2~meV. Since this process is the only mechanism of deceleration at
the lowest energies, the mean energy of muonic atoms at $T\lesssim{}5$~K
is much greater than the corresponding energy in a~perfect-gas target.
Also, the deceleration process below 10~meV becomes much slower at such
temperatures, since the elastic Bragg scattering in a~heavy target
cannot change the kinetic energy of the muonic atoms.

Above about 10 meV, the rotational and then vibrational excitations of
the target molecules support a~very effective mechanism for the
deceleration of muonic atom. At $\varepsilon\gtrsim{}100$~meV, the cross
sections for the hydrogenic crystals (per one molecule) tend to those
calculated for the isolated hydrogenic molecules. However, the effective
temperature of molecules in the hydrogenic crystals is much higher than
the target temperature, due to the zero-point vibrations of the
molecules in the lattice. This effect does not disappear (even at high
collision energies).

Although coherent and incoherent phonon effects have been estimated in
the simplest approach of the isotropic Debye solid, the Monte Carlo
simulations using the calculated differential cross sections are in good
agreement with the available experimental data.


\bibliography{adamczak}

\end{document}